\documentclass[12pt,a4paper,fleqn]{article}
\usepackage[textheight=23cm,textwidth=16cm]{geometry}
\usepackage{amsmath}
\usepackage{amssymb}
\usepackage{graphicx}
\usepackage{cite}
\usepackage{here}
\usepackage[american]{babel}

\usepackage{tabularx}

\begin{document}

\begin{center}

{\LARGE\bf
 Gradient-Driven Molecule Construction: \\
An Inverse Approach Applied to the Design of Small-Molecule Fixating Catalysts
}

\vspace{0.2cm}

{\large
Thomas Weymuth
and Markus Reiher\footnote{E-Mail: markus.reiher@phys.chem.ethz.ch}
}\\[2ex]

ETH Zurich, Laboratorium f\"ur Physikalische Chemie, \\
Vladimir-Prelog-Weg 2, 8093 Zurich, Switzerland

%\vspace{1cm}

\end{center}

\begin{abstract}

Rational design of molecules and materials usually requires extensive screening of molecular structures for the desired property. 
The inverse approach to deduce a structure for a predefined property would be highly desirable, but is, unfortunately, not 
well-defined. However, feasible strategies for such an inverse design process may be successfully developed for specific purposes. 
We discuss options for calculating ``jacket'' potentials that fulfill a predefined target requirement\,---\,a concept that we 
recently introduced [T.~Weymuth, M.~Reiher, {\it MRS Proceediungs} {\bf 2013}, {\it 1524}, DOI:10.1557/opl.2012.1764]. We 
consider the case of small-molecule activating transition metal catalysts. As a target requirement we choose the vanishing 
geometry gradients on all atoms of a subsystem consisting of a metal center binding the small molecule to be activated. The 
jacket potential can be represented within a full quantum model or by a sequence of approximations of which a field of electrostatic 
point charges is the simplest. In a second step, the jacket potential needs to be replaced by a chemically viable chelate-ligand 
structure for which the geometry gradients on all of its atoms are also required to vanish. In order to analyze the feasibility 
of this approach, we dissect a known dinitrogen-fixating catalyst to study possible design strategies that must eventually produce 
the known catalyst.

\end{abstract}

\begin{tabbing}
Date:   \quad \= March 24, 2014 \\
Status:       \> printed in {\it Int.~J.~Quantum Chem.}, {\it 114} {\bf 2014} 838--850 \\
\end{tabbing}

%\newpage
\protect
\section{Introduction}
\label{sec:intro}

The design of new molecules and materials exhibiting favourable properties is an ever ongoing quest in a wide range of research 
fields. With the advent of powerful computer systems, theoretical methods and tools are becoming increasingly important in this 
design process\cite{nature_1999_402_60,acc_chem_res_2006_39_71}. These methods allow one to pre-screen properties of a set of 
molecules or materials without the need of time- and resource-intensive synthesis. 

In these so-called direct methods it is necessary to first specify the structural composition of the compound to be studied. 
However, the structure is usually not exactly known\,---\,in fact, the precise knowledge of it is the goal of the design process. Therefore, 
an approach which finds a structural composition compatible with a predefined property would be highly desirable. Such approaches 
are usually called inverse approaches. Although some pioneering work on inverse quantum chemical approaches has already been carried 
out (see Refs.~\cite{sci_china_ser_b_chem_2009_52_1769,int_j_quantum_chem_2009_109_2456,int_j_quantum_chem_doi101002qua24375,
chimia_2009_63_270,acc_chem_res_2006_39_71,weym14b} for reviews, the search for a structure which 
features a given predefined property remains an extremely complex computational challenge because of the sheer size of chemical 
compound space, i.e., of the set of all compounds accessible with contemporary synthetic protocols.

A problem-specific development of inverse quantum chemical methods could be more beneficial. Here, we develop an approach for 
the design of molecules and materials with specific stability. We shall first define the general principle of our inverse 
approach and then illustrate implementations at a specific example.

Rational design and thus also inverse approaches should rely on the free energy as the ultimate thermodynamic criterion to 
decide on the stability of a computationally designed molecule or material. However, for reactions which involve the formation 
of strong bonds, the reaction free energy is determined mostly by the difference in electronic energy of the reactants. In 
such cases, electronic structure theory at zero Kelvin is usually sufficient for the design process and temperature corrections 
may be neglected. Then, the minimum of the electronic energy of a product molecule or material can serve as a stability criterion. 
This minimum is determined by its vanishing geometry gradients (under the assumption that all eigenvalues of the geometry Hessian 
are positive). Hence, following our previous initial investigation\cite{mrs_proceedings_2013_1524_doi} we may formulate the 
design principle as follows: 

\vspace{0.5cm}
\begin{center}
\framebox{
\begin{minipage}{14cm}
{\bf Gradient-driven Molecule Construction (GdMC):} Assume that a desired molecular property or function can be expressed in terms of a certain
(possibly idealized) structural feature. This feature considered as an isolated fragment is as such in general not stable, i.e., it features non-vanishing 
geometry-gradient components (forces) on its atoms.
If, however, we consider this fragment as a subsystem of a larger, unknown target molecule or material, then
its unknown parts can be designed such that the geometry gradient of the electronic energy on all nuclei of the full target molecule or material 
(i.e., including the pre-defined fragment and the (to be) constructed complementary subsystem) approaches zero component-wise.
The overall structural composition of the target molecule or material may then be determined under this constraint. 
\end{minipage}
}
\end{center}
\vspace{0.5cm}

An example for the application of this design principle is the search for small-molecule activating metal complexes. Important 
small molecules, for which activation procedures are desirable, are H$_2$, N$_2$, O$_2$, CO$_2$, and CH$_4$. Upon fixation of 
such a small molecule by a transition metal ion embedded in a surface or in a sophisticated chelate ligand, a strong bond can be 
formed, which weakens the bonds within the small molecule and may induce other structural changes (like bending due to an evolving 
lone pair).

\begin{figure}[H]
 \centering
 \includegraphics{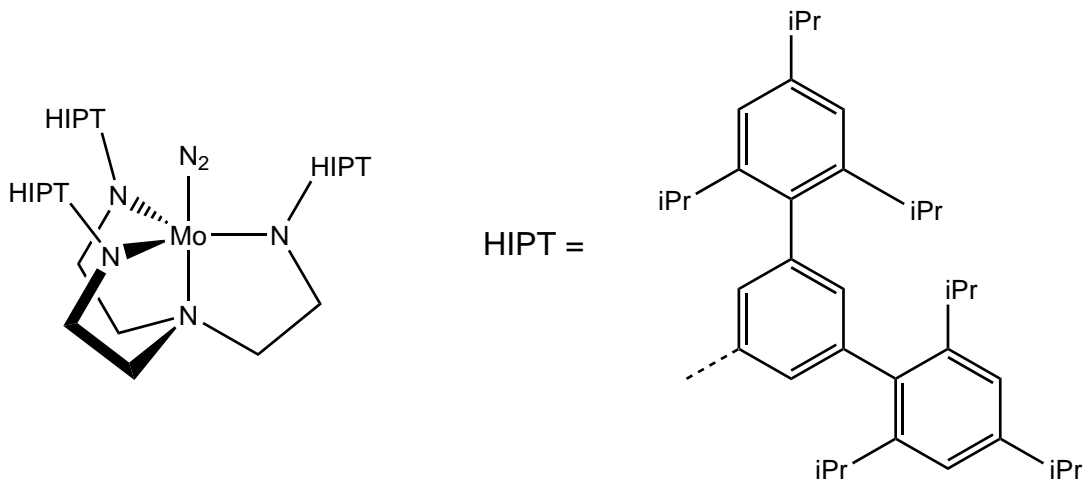}
 \caption{Lewis structure of the original Schrock catalyst, [MoN(NC$_2$H$_2$HIPT)$_3$)]N$_2$ (HIPT: hexa-{\it iso}-propyl terphenyl), 
the first homogeneous catalyst capable of activating dinitrogen and transforming it into ammonia under ambient conditions \cite{science_2003_301_76}.}
 \label{fig:schrock}
\end{figure}

A specific example is the task of finding a transition metal catalyst that reduces molecular dinitrogen to ammonia under ambient 
conditions at high turnover numbers\cite{angew_chem_int_ed_2006_45_6264,acc_chem_res_2005_38_955,coord_chem_rev_2013_257_587,
nature_chem_2013_5_559,acc_chem_res_1997_30_460}. In industry, the process is accomplished by a heterogeneous iron catalyst within 
the Haber--Bosch process, but under harsh conditions\cite{leig02}. Only three different families of synthetic homogeneous catalysts, 
two relying on molybdenum and one on iron as central transition metal atom, have achieved this task so far\cite{science_2003_301_76,j_am_chem_soc_2004_126_6150,nature_chem_2011_3_120,organometallics_2012_31_8437,nature_2013_501_84}, 
but they are in general all suffering from low turn-over numbers. Intense research efforts\cite{proc_nat_acad_sci_2006_103_17099,nature_chem_2011_3_95,angew_chem_int_ed_2008_47_5512,inorg_chem_2009_48_1638} focused 
on increasing the stability of one of them, the Schrock catalyst depicted in Fig.~\ref{fig:schrock}.

With quantum chemical methods, we were able to identify\cite{inorg_chem_2009_48_1638,chem_eur_j_2009_15_5073,inorg_chem_2008_47_3634,angew_chem_int_ed_2006_45_6264,inorg_chem_2005_44_9640,
chem_eur_j_2005_11_7448,chem_eur_j_2005_11_574,chem_eur_j_2004_10_4443,chem_eur_j_2004_10_819,chem_eur_j_2004_10_4214,adv_inorg_chem_2004_56_55,
inorg_chem_acta_2003_348_194,chem_eur_j_2002_8_5332} the most crucial steps in catalytic processes that facilitate N$_2$ fixation: 1), 
feasibility of dinitrogen binding, 2), transfer of a first hydrogen atom (e.g., as separate proton and electron transfers) onto the 
bound dinitrogen ligand, 3), exchange of the (second) ammonia molecule produced by the next incoming dinitrogen ligand, and 4), 
prevention of side reactions.

For a nitrogen-fixating catalyst, the first step is thus to bind dinitrogen to one, two or even several transition metal centers 
in some ligand environment. This task may be very difficult to achieve, as is highlighted, for instance, by the case of iron 
complexes in a sulfur-rich ligand environment\cite{adv_inorg_chem_2004_56_55,chem_eur_j_2002_8_5332,acc_chem_res_1997_30_460,
coord_chem_rev_1999_190_607,coord_chem_rev_2000_200_545} (note that the active site of nitrogenase is an iron-sulfur 
cluster\cite{science_1992_257_1677}). These iron complexes bind dinitrogen rather weakly and feature hardly any bond activation 
as monitored by the N--N bond length that remains almost unchanged. However, an N$_2$-fixating system may directly activate 
N$_2$ upon binding such that the triple bond is broken and a diazenoid or even a hydrazinoid structure with corresponding 
elongated N--N bond length emerges\cite{coord_chem_rev_2010_254_1883}. If the N--N bond is not activated upon coordination, a 
subsequent reduction by electron transfer onto the ligand is necessary in order to start the chemical reduction 
process\cite{chem_eur_j_2005_11_574,chem_eur_j_2004_10_4443}.

In a first application of the GdMC design principle to the task of identifying a dinitrogen-binding or dinitrogen-activating 
complex, we started\cite{mrs_proceedings_2013_1524_doi} from a predefined central fragment, which consists of a molybenum 
metal atom and N$_2$ at a certain distance with the N--N bond length fixed to some reasonable value. The task is then to find 
a ligand environment that reduces the geometry gradients on all atoms in the compound. While we have studied already some 
options and possibilities for Schrock-type dinitrogen fixation within this framework\cite{mrs_proceedings_2013_1524_doi}, a 
rigorous investigation is mandatory and shall be provided in this work. Therefore, we first consider the basic formalism of 
GdMC in the subsequent section and discuss possible realizations of gradient-reducing potentials to be substituted by a fragment 
scaffold in a second step. The whole analysis in the subsequent results section is guided by the knowledge that we have gained 
in the past decade about the Schrock dinitrogen activating molybdenum complex. The idea is to understand what measures need to 
be taken in order to reconstruct a complex whose functionality has already been confirmed and investigated.

\section{Theory of Gradient-Driven Molecule Construction}

The definition of a molecular fragment structure that shall be stabilized by chemical embedding into a surrounding molecular 
scaffold is central to GdMC. Without the chemical embedding, the fragment structure will in general not represent a stationary 
point on the potential energy surface. In other words, without the chemical embedding the fragment will feature a non-vanishing 
geometry gradient on the nuclei of the fragment (i.e., the derivative of the total electronic energy with respect to the nuclear 
coordinates will be significantly different from zero).  Moreover, the GdMC concept requires to construct a molecular environment 
such that the {\it overall} gradient will vanish and the resulting compound will represent a stable structure on the electronic 
energy hypersurface. 

For an algorithmic realization of this concept, atomic nuclei and electrons must be placed to form a suitable scaffold that can 
host and stabilize the fragment. Accordingly, the number of electrons as well as the number, position, and charge of atomic 
nuclei are optimization parameters. This optimization problem is very complex (even if we neglect for the moment the spin 
degrees of freedom). Not only the lengths of all Cartesian gradient components must vanish at each nucleus of the fragment, 
they must also vanish at all nuclei which are added in the chemical embedding process. However, this second requirement can 
be separated from the first one if a two-step optimization procedure is adopted. For this, the gradient-reducing environment 
may be represented by a jacket potential $\upsilon_{\rm jac}$ that mediates all interactions with the fragment\cite{mrs_proceedings_2013_1524_doi}. 
In a second step, a molecular realization of the optimized jacket potential needs to be found. It is the purpose of this paper 
to elaborate on possible design strategies in such one- and two-step approaches.

We choose the jacket potential $\upsilon_{\rm jac}$ to enter the electronic fragment Hamiltonian $H_{\rm el}$ (in Hartree 
atomic units) as a one-electron operator,
\begin{equation}\label{eq:h}
 H_{\rm el} = - \frac{1}{2}\sum_{i\in{\rm frag}}\Delta_i - \sum_{i,I\in{\rm frag}}\frac{Z_I}{r_{iI}} + \sum_{i,j>i \in {\rm frag}}\frac{1}{r_{ij}} + \sum_{I,J>I \in {\rm frag}}\frac{Z_IZ_J}{r_{IJ}} + \sum_{i \in {\rm frag}} \upsilon_{\rm jac}(i).
\end{equation}
where the indices $I$, $J$ and $i$, $j$ run over all nuclei and electrons, respectively, of the fragment. 
$Z_I$ is the charge number of nucleus $I$, and $r_{ij}$  is the spatial distance between particles $i$ and $j$
(the fragment is denoted as ``frag''). The first term 
describes the kinetic energy of fragment electrons with the Laplacian $\Delta_i$. Although the jacket potential is written 
as a one-electron operator, it may contain contributions from nuclear repulsion terms, electronic kinetic energy operators and 
so forth when it is replaced by a viable molecular structure in the second optimization step. 

In fact, the particular choice of the jacket potential will determine the approximation adopted in the second step. If a purely 
classical (electrostatic) embedding is considered to be adequate for the design problem (i.e., if exchange and quantum correlation 
effects can be neglected), then the jacket potential will exactly be a one-electron operator. If, however, quantum mechanical 
superposition effects are non-negligible\,---\,as this would be the case for a fragment that requires cuts through chemical bonds 
to separate them from the environment\,---\,then two-electron effects or nonadditive kinetic energy terms could become important. 

The total electronic energy can be obtained as the expectation value of the above Hamiltonian,
\begin{equation}
 E_{\rm el} = \langle\Psi_{\rm el}|H_{\rm el}|\Psi_{\rm el}\rangle.
\end{equation}

If we approximate the many-electron wave function $\Psi_{\rm el}$ of the subsystem by a determinant expansion, the orbitals 
can be obtained from a self-consistent-field-type equation,
\begin{equation}\label{eq:addpot}
 \left( \frac{1}{2}\gamma_{ii}\Delta + \upsilon_{\rm frag}(\boldsymbol{r}) + \upsilon_{\rm jac}(\boldsymbol{r}) \right)\phi_i(\boldsymbol{r}) = \varepsilon_i\phi_i(\boldsymbol{r}),
\end{equation}
where $\phi_i(\boldsymbol{r})$ and $\varepsilon_i$ represent the $i$-th orbital and the associated orbital energy of the fragment, respecticely.
$\gamma_{ii}$ is a generalized occupation number, which is equal to one for the spin orbitals of a 
single-determinant theory like Hartree--Fock or Kohn--Sham density functional theory (DFT) and which is a real number in multi-determinant 
theories. The first term of the operator on the left-hand side represents the kinetic energy operator for an electron (of the fragment), and the 
second term collects all potential energy terms important for the fragment,
\begin{equation}\label{eq:syspot}
  \upsilon_{\rm  frag}(\boldsymbol{r}) =  \upsilon_{\rm  frag}^{\rm  (ne)}(\boldsymbol{r}) + \upsilon_{\rm frag}^{\rm (coul)}(\boldsymbol{r}) + \upsilon_{\rm frag}^{\rm (xc)}(\boldsymbol{r}) + \upsilon_{\rm frag}^{\rm (nn)}(\boldsymbol{r}),
\end{equation}
in which the individual terms denote the attraction of nuclei (n) and electrons (e), the pairwise repulsion of all electrons (coul), 
a non-classical term representing exchange--correlation (xc) effects and finally the nucleus--nucleus repulsion within the fragment. 

The third term on the left-hand side of Eq.~(\ref{eq:addpot}) is the jacket potential. It represents both the energy contribution 
of the ligand sphere (to which we will refer as ``the environment'') as well as its interaction with the fragment. We may divide 
it into a potential energy operator for the environment and one for the interaction between the fragment and the ligand environment,
\begin{equation}
 \upsilon_{\rm jac}(\boldsymbol{r}) = \upsilon_{\rm env}(\boldsymbol{r}) + \upsilon_{\rm int}(\boldsymbol{r}).
\end{equation}
Formally, we may write the environment potential as
\begin{equation}
 \upsilon_{\rm env}(\boldsymbol{r}) = \upsilon_{\rm env}^{\rm (ne)}(\boldsymbol{r}) + \upsilon_{\rm env}^{\rm (coul)}(\boldsymbol{r}) + \upsilon_{\rm env}^{\rm (xc)}(\boldsymbol{r}) + \upsilon_{\rm env}^{\rm (nn)}(\boldsymbol{r}) + \upsilon_{\rm env}^{\rm (kin)}(\boldsymbol{r}),
\end{equation}
where the first three terms have a similar meaning as in Eq.~(\ref{eq:syspot}) and where the fifth term collects the kinetic 
energy contribution of environment electrons.  Analogously, we may split the interaction part into five contributions,
\begin{equation}\label{eq:intpot}
 \upsilon_{\rm int}(\boldsymbol{r}) = \upsilon_{\rm int}^{\rm (ne)}(\boldsymbol{r}) + \upsilon_{\rm int}^{\rm (coul)}(\boldsymbol{r}) + \upsilon_{\rm int}^{\rm (xc)}(\boldsymbol{r}) + \upsilon_{\rm int}^{\rm (nn)}(\boldsymbol{r}) + \upsilon_{\rm int}^{\rm (kin)}(\boldsymbol{r}).
\end{equation}
In this equation, the first term represents the interaction of nuclei of the fragment with electrons of the environment
and the interaction between nuclei of the environment and electrons of the fragment. The second term stands for the 
Coulomb repulsion of electrons of the fragment and the environment while the third term and the last term represent the  
exchange--correlation contribution from the fragment and the ligand sphere and a possible non-additive kinetic-energy contribution 
for the joined fragment and environment, respectively. The fourth term in Eq.~(\ref{eq:intpot}) denotes the mutual repulsion 
between nuclei of the fragment and those of the environment.

Before we continue, we note that a jacket-potential construction may lead to a sequence of one-electron equations which all
represent the electronic and geometric structure of a molecular system that eventually will become identical to the target system,
when all geometry-gradient components vanish (see next paragraph). Interestingly, each set of one-electron equations corresponds to
an underlying many-body system, which could be reconstructed. Such a 'reverse engineering' approach has recently
been discussed for a potential improvement of density functionals, for which a spatially resolved error could be exploited \cite{coe-09}
(see also Ref.\ \cite{boguslawski13} and references cited therein). Note. however, that such 'reverse engineering' approaches are
more involved than the C-representability problem \cite{j_am_chem_soc_2006_128_3228} we are aiming at, namely one that will produce a nuclear 
framework and global 'quantum numbers' (number of electrons, total charge, global spin state) of the target molecule.

We can now define the absolute value of the geometry gradient of one nucleus $I$ as 
\begin{equation}\label{eq:abgrad}
 |\nabla_I E_{\rm el}| = \sqrt{\left( \frac{\partial E_{\rm el}}{\partial r_{I, x}} \right)^2 + \left( \frac{\partial E_{\rm el}}{\partial r_{I, y}} \right)^2 + \left( \frac{\partial E_{\rm el}}{\partial r_{I, z}} \right)^2},
\end{equation}
For the analysis to come, we define an overall absolute gradient $|\nabla_{\rm frag}E_{\rm el}|$ of a fragment as the sum of 
all individual absolute gradients of that fragment,
\begin{equation}
 |\nabla_{\rm frag}E_{\rm el}| = \sum_{B \in {\rm frag}}|\nabla_B E_{\rm el}|.
\end{equation}
The individual terms in the square root expression of Eq.~(\ref{eq:abgrad}) are given by
\begin{align}\label{eq:grad}
 \frac{\partial E_{\rm el}}{\partial r_{I,\alpha}} = & \sum_{B \in {\rm frag}} \frac{Z_I Z_B (r_{I,\alpha} - r_{B,\alpha})}{|\boldsymbol{r}_{I} - \boldsymbol{r}_{B}|^3} - \sum_i\gamma_{ii}\left\langle\left.\left.\frac{\partial \phi_i(\boldsymbol{r})}{\partial r_{I, \alpha}}\right|\Delta\right|\phi_i(\boldsymbol{r})\right\rangle \\ \nonumber
  & + \int\frac{\partial \upsilon_{\rm frag}(\boldsymbol{r})}{\partial r_{I, \alpha}}\rho(\boldsymbol{r}){\rm d}\boldsymbol{r} + \int\upsilon_{\rm frag}(\boldsymbol{r})\frac{\partial \rho(\boldsymbol{r})}{\partial r_{I, \alpha}}{\rm d}\boldsymbol{r} \\ \nonumber
  & + \int\frac{\partial \upsilon_{\rm jac}(\boldsymbol{r})}{\partial r_{I, \alpha}}\rho(\boldsymbol{r}){\rm d}\boldsymbol{r} + \int\upsilon_{\rm jac}(\boldsymbol{r})\frac{\partial \rho(\boldsymbol{r})}{\partial r_{I, \alpha}}{\rm d}\boldsymbol{r},
\end{align}
where $\alpha \in \{x, y, z\}$.
In Eq.~(\ref{eq:grad}), the first two terms are derivatives of the nucleus--nucleus repulsion energy and the electronic 
kinetic energy of the fragment. The remaining terms are derivatives of the potentials occurring in Eq.~(\ref{eq:addpot}); 
$\rho(\boldsymbol{r})$ is the total electron density of fragment and environment and $\boldsymbol{r}$ a spatial electron 
coordinate. Of course, the potential terms can be broken down into derivatives of their individual contributions given in 
Eqs.~(\ref{eq:syspot})\,--\,(\ref{eq:intpot}). In particular, we have
\begin{equation}
 \frac{\partial \upsilon_{\rm frag}^{\rm (ne)}(\boldsymbol{r})}{\partial r_{I, \alpha}} = \frac{-Z_I (r_{I,\alpha} - r_{\alpha})}{|\boldsymbol{r}_{I} - \boldsymbol{r}|^3},
\end{equation}
%Das folgende ist falsch, da es eine Energie und kein Potential ist:
%and
%\begin{equation}\label{eq:coulgrad}
%  \frac{\partial \upsilon_{\rm frag}^{\rm (coul)}(\boldsymbol{r})}{\partial r_{I, \alpha}} = \int\int \frac{\partial \rho(\boldsymbol{r})}{\partial r_{I, \alpha}} \frac{\rho(\boldsymbol{r}')}{|\boldsymbol{r} - \boldsymbol{r}'|}{\rm d}\boldsymbol{r}{\rm d}\boldsymbol{r}'.
%\end{equation}
All other derivatives are not straightforward to evaluate, because their explicit form has deliberately not been specified 
in this general formalism. It depends on the approximations made in a specific exchange--correlation functional as well as on 
the environment and interaction potentials. However, from Eq.~(\ref{eq:grad}) we understand that an 
important quantity in the gradient expression is the derivative of the electron density $\rho(\boldsymbol{r})$ with respect 
to the nuclear coordinates. 

For the sake of simplicity, we carry out the derivations for the simple case of a one-determinant approximation to $\Psi_{\rm el}$ 
as employed in Hartree--Fock theory or in Kohn--Sham DFT. In this case, the density is a sum over the absolute squares of all 
occupied orbitals,
\begin{equation}
 \rho(\boldsymbol{r}) = \sum_i |\phi_i(\boldsymbol{r})|^2,
\end{equation}
from which we obtain for real orbitals
\begin{equation}\label{eq:drho}
 \frac{\partial \rho(\boldsymbol{r})}{\partial r_{I, \alpha}} = 2\sum_i \phi_i(\boldsymbol{r}) \frac{\partial \phi_i(\boldsymbol{r})}{\partial r_{I, \alpha}}.
\end{equation}
In molecular calculations, the orbitals are usually expanded in terms of atom-centered basis functions,
\begin{equation}
 \phi_i(\boldsymbol{r}) = \sum_{\mu}\sum_{B} c_{\mu i}\chi_{\mu B}(\boldsymbol{r} - \boldsymbol{r}_B),
\end{equation}
so that we find
\begin{equation}\label{eq:gradphi}
 \frac{\partial \phi_i(\boldsymbol{r})}{\partial r_{I, \alpha}} = \sum_{\mu} \frac{\partial c_{\mu i}}{\partial r_{I, \alpha}} \chi_{\mu I}(\boldsymbol{r} - \boldsymbol{r}_I) + \sum_{\mu} c_{\mu i} \frac{\partial \chi_{\mu I}(\boldsymbol{r} - \boldsymbol{r}_I)}{\partial r_{I, \alpha}},
\end{equation}
leads to the so-called Pulay forces\cite{pula87,reih09}.
For the known basis functions $\chi_{\mu I}$ their derivative with respect to $r_{I, \alpha}$ is straightforward to evaluate.

Our GdMC concept requires that all geometry gradients vanish, i.e.,
\begin{equation}
 |\nabla_I E_{\rm el}| \stackrel{!}{=} 0 \hspace{1cm} \forall I,
\end{equation}
which is only possible if {\it all} individual Cartesian components vanish, as can be seen by Eq.~(\ref{eq:abgrad}). Thus, 
from requiring that the right hand side of Eq.~(\ref{eq:grad}) is zero, we obtain
\begin{align}\label{eq:grad0}
 & -\sum_B \frac{Z_I Z_B (r_{I,\alpha} - r_{B,\alpha})}{|\boldsymbol{r}_{I} - \boldsymbol{r}_{B}|^3} + \sum_i\gamma_{ii}\left\langle\frac{\partial \phi_i(\boldsymbol{r})}{\partial r_{I, \alpha}}\left|\Delta\right|\phi_i(\boldsymbol{r})\right\rangle \\ \nonumber
 & - \int\frac{\partial \upsilon_{\rm frag}(\boldsymbol{r})}{\partial r_{I, \alpha}}\rho(\boldsymbol{r}){\rm d}\boldsymbol{r} - \int\upsilon_{\rm frag}(\boldsymbol{r})\frac{\partial \rho(\boldsymbol{r})}{\partial r_{I, \alpha}}{\rm d}\boldsymbol{r} \nonumber \\
 & \stackrel{!}{=} \int\frac{\partial \upsilon_{\rm jac}(\boldsymbol{r})}{\partial r_{I, \alpha}}\rho(\boldsymbol{r}){\rm d}\boldsymbol{r} + \int\upsilon_{\rm jac}(\boldsymbol{r})\frac{\partial \rho(\boldsymbol{r})}{\partial r_{I, \alpha}}{\rm d}\boldsymbol{r} \hspace{1cm} \forall I {\rm ~and~} \alpha.
\end{align}
This equation might be used as a working equation to determine $\upsilon_{\rm jac}(\boldsymbol{r})$. Its solution is certainly 
not trivial, but it might be approximated with iterative numerical methods. In order to understand how this can be achieved, we 
take a step back and analyze the construction of a chemical environment at the specific example of the Schrock dinitrogen-fixation complex.

\section{Computational Details}

All calculations presented in this work have been carried out within the density functional theory framework. 
The results in Section \ref{sec:potential} were obtained with a local version of {\sc Turbomole} 5.7.1, which was modified 
such that the jacket potential can be represented on the DFT grid. In all {\sc Turbomole} calculations, we applied the BP86 
exchange--correlation functional\cite{phys_rev_a_1988_38_3098,phys_rev_b_1986_33_8822}, in combination with Ahlrichs' def-TZVP 
Gaussian-type basis set at all atoms\cite{j_chem_phys_1994_100_5829}. Note that the (standard) resolution-of-the-identity 
approximation was \textit{not} invoked. Stuttgart effective core potentials were applied to Mo as implemented in {\sc Turbomole}.

For the calculations in the remaining sections, we employed the program package {\sc Adf}, version 2010.02b\cite{j_comput_chem_2001_22_931}, 
with the BP86 exchange--correlation functional\cite{phys_rev_a_1988_38_3098,phys_rev_b_1986_33_8822} and the Slater-type TZP 
basis set without frozen cores\cite{j_comput_chem_2003_24_1142} at all atoms. In these calculations advantage was taken of the 
resolution-of-the-identity technique for the evaluation of the two-electron Coulomb integrals. Scalar-relativistic effects were 
taken into account for all atoms by means of the zeroth-order regular approximation (ZORA)\cite{j_chem_phys_1993_99_4597}. 

We should explicitly state here that the actual numerical values for nuclear gradients depend on the basis set used. Therefore, 
it is important that such gradients are calculated using exactly the same procedure in order to be able to compare them to each 
other.

\section{Model Hierarchies}

The jacket potential representing the environment in GdMC and its interaction with the fragment can be modeled with a range of 
different approaches. As already stated above, the exact mathematical form of the environment and interaction potentials depends 
strongly on the ansatz chosen for the approximation of the electronic wave function. In a formally exact treatment, we take the 
ligand sphere as being constituted by nuclei and electrons, but we do not know their exact nature, number, and spatial distribution 
prior to the GdMC optimization. Moreover, for the accurate calculation of all contributions to the total potential energy, we 
would require a framework that can deal with the quantum basis states on fragment and environment in a theory of open quantum 
systems\cite{aman11}. Even though a simplified, although formally still exact, frozen-density embedding framework\cite{j_phys_chem_1993_97_8050,phys_rep_2010_489_1,annu_rep_prog_chem_sect_c_phys_chem_2012_108_222} 
would simplify this issue significantly (leaving aside the fact that a scission of covalent or dative bonds can introduce nonnegligible
errors \cite{kiew08,fux-08,fux-10}), the optimization problem is still highly involved. Hence, the terms in the jacket 
potential need to be approximated in order to arrive at a computationally feasible approach. In the sections to follow, we will 
investigate three different approaches.

\subsection{Environment Potential}
\label{sec:potential}

Within the framework of DFT a grid of mesh points in position space is usually employed for the numerical integration of the 
exchange--correlation energy (and its functional derivative). It is therefore easy to represent a jacket potential on this grid 
and to introduce it into the Kohn--Sham equations\,---\,provided that the thinning of mesh points in the exponentially decaying 
asymptotic region of the fragment electron density is stalled to account for points in the environment to be constructed. With 
established optimization algorithms (simplex, simulated annealing, genetic algorithms, etc.) it should be possible to optimize 
an additional potential such that the geometry gradient is as small as possible. 

As the grid is tailored to represent the potential in the close vicinity to the central fragment, the grid centered around this 
central fragment will most likely not extend far enough into space to allow for the accurate representation of a large chelate 
ligand sphere. While a tailored approach for extending the thin DFT integration grids asymptotically is certainly feasible, one 
could also aim for a stepwise build-up of the ligand sphere by first reconstructing the ligand atoms directly connected to the 
central fragment and then performing additional sequential optimizations until the gradient is sufficiently small. 

Of course, it can be anticipated that the optimization strongly depends on the starting guess. A global optimization is very likely 
not to be feasible, since even very coarse DFT grids for small model system have a rather large number of points. For example, 
for a transition metal catalyst suitable for nitrogen fixation, a Mo--N$_2$ fragment appears suitable as central fragment (cf., 
Fig.~\ref{fig:schrock}; in fact, we will use this very fragment in the later sections) The coarsest grid in {\sc Turbomole} (grid ''1``) 
has almost 7000 points for the Mo--N$_2$ fragment and, thus, 7000 parameters would have to be optimized. Moreover, the reconstruction 
of an actual molecular ligand sphere from such an abstract potential will not be obvious unless additional constraints are introduced 
to ensure that a solution is optimized which is actually representable by a chemical structure (i.e., that the potential is C-representable).

\begin{figure}[H]
 \centering
 \includegraphics[scale=0.2]{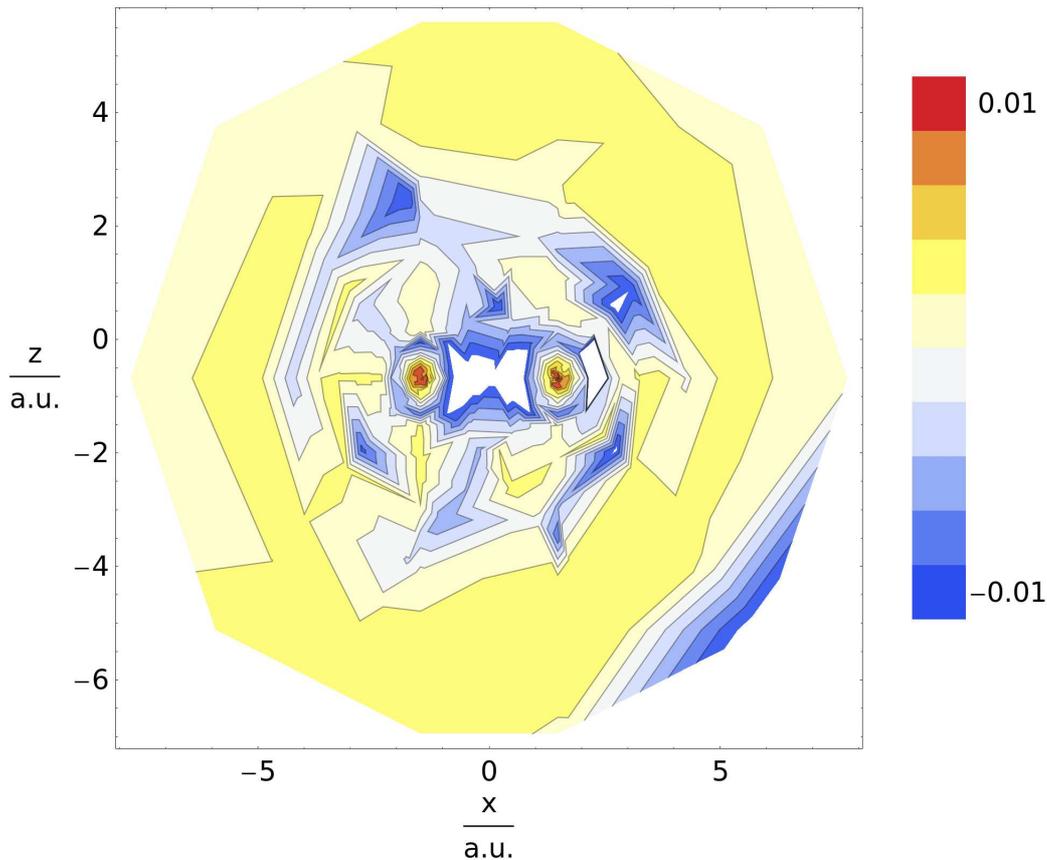}
 \caption{Contour plot of the jacket potential in the xz-plane, in which the H$\cdots$H fragment is oriented. The positions of the two hydrogen
nuclei are where the jacket potential takes highest values. The oxygen nucleus is in the origin of the coordinate system. White regions indicate
values of the potential beyond the scale chosen. All data are given in atomic units (a.u.).}
 \label{fig:potential}
\end{figure}

Nevertheless, the approach is appealing and we shall therefore report our preliminary investigations here. We modified {\sc Turbomole} 
5.7.1 such that an arbitrary potential $\upsilon_{\rm jac}$ can be read in, which is then added to the Kohn--Sham potential. Then, 
we used the simplex algorithm in order to minimize the overall gradient on the central fragment. As a starting guess for the 
additional potential, we utilized a zero vector, as this appears to be the most natural choice without relying on any assumption. 
In order to keep the optimization problem as simple as possible, we chose the optimized structure of a water molecule and removed 
the oxygen atom, thereby creating a dihydrogen molecule with a signficantly enlarged bond. The goal is now to find a potential 
stabilizing this H$\cdots$H fragment. In this case, the variational problem involved 1706 parameters. It was possible to find a 
potential reducing the gradient from 1.50\textperiodcentered10$^{-1}$\,hartree/bohr to 8.66\textperiodcentered10$^{-3}$\,hartree/bohr, 
which clearly demonstrates that no principle obstacles exist for this approach. The stabilizing potential is depicted in Fig.~\ref{fig:potential}.

From Fig.~\ref{fig:potential} we understand that the potential features maxima at the positions of the two hydrogen nuclei. There are some interesting 
features, most notably we identify local minima in the vincinity of the nuclei. Certainly, a ligand sphere cannot be easily constructed from this 
potential. One reason for this is that the structure of the optimized potential in Fig.~\ref{fig:potential} shows that there exists no unique
solution to the optimization problem. On the one hand, the potential does not even obey $C_{2v}$ point-group symmetry. On the other hand, it is
obvious that additional features of the potential --- like an analytic short- and long-range behavior --- should be enforced in the optimization
procedure. Moreover, Wang {\it et al.}~showed how to solve this problem of C-representability in a very elegant fashion, namely by 
expanding the external potential into a set of atomic potentials\cite{j_am_chem_soc_2006_128_3228}. Still, the implementation of 
this method does not solve the problem that even for comparatively small central fragments and very coarse DFT grids, the number 
of grid points and therefore parameters to be optimized remains very large. Also efficient optimization algorithms, combined with 
a parallel evaluation of the geometry gradient on fast contemporary computer hardware, need several days to weeks (possibly even 
months) to find a solution. It is obvious that this approach can only be turned into an efficient method if additional analytic 
constraints on the optimized potential can be included (e.g., by exploiting analytic knowledge about the potential itself or about the electron density like the 
asymptotically exponential decay or the nuclear cusp condition; cf.\ Ref.\ \cite{neug02}).

\subsection{Environment Potential Represented by Point Charges}
\label{sec:pointcharges}

In order to achieve general applicability, we are advised to establish an abstract optimization scenario, in which the ligand sphere 
is represented by an abstract interaction potential, which in a first approximation may simply be represented by a collection of 
(fractional) point charges to be placed anywhere in space\,---\,not necessarily located at the position of an atomic nucleus. With 
enough such charges, we should be able to represent a charge density that produces an electrostatic potential via Poisson's equation 
(neglecting in a first step quantum superposition (entanglement) effects, i.e., $\upsilon_{\rm env}^{\rm (xc)} = 0$ and $\upsilon_{\rm int}^{\rm (xc)} = 0$). 
From this charge density, the corresponding ligand sphere could be deduced provided that a solution is optimized which obeys the 
typical characteristics of electronic plus nuclear charge densities. An advantage of this approach is that it is straightforward to 
implement as most quantum chemical programs can deal with arbitrary fields of point charges although present self-consistent field 
convergence accelerators may not be optimal to converge orbitals in such fields. Still, the resulting optimization problem is very 
hard. We have to optimize the number of point charges, as well as their spatial distribution and the values of the individual charges.

In a first investigation, we started with few point charges in order to keep the variational problem feasible. It is necessary to 
employ a global optimization algorithm, for which we implemented a simple interface between {\sc Mathematica}\cite{mathematica9} 
and {\sc Adf}\cite{j_comput_chem_2001_22_931} in order to exploit the sophisticated optimization routines implemented in {\sc Mathematica}. 
This setup allows us to carry out global optimizations of the overall gradient with methods such as simulated annealing, random walks, 
and differential evolution\cite{j_global_optim_1997_11_341,ieee_trans_evol_comp_2011_15_4}. We focus on the latter algorithm. 

Following the recommendations given in Ref.~\cite{j_global_optim_1997_11_341}, we employed a scaling factor $s$ of 0.5, a 
cross probability of 0.9, and a population size ten times the number of optimization parameters. In all optimizations, a 
random seed of zero was used. The solution was assumed to be converged when two subsequent best function values differed by 
less than 10$^{-5}$\,hartree/bohr and the two best solution vectors differed by less than 10$^{-3}$\,hartree/bohr. In a first 
study, we optimized the position and magnitude of a single point charge, i.e., the number of optimization parameters was only 
four. As fragment the central part of the Schrock catalyst, i.e., the triatomic moiety Mo--N$_2$, was chosen. The single 
point charge was confined to be between $-$8 and 8\,bohr in each of the three coordinates (which encompasses the central fragment
at positions (in bohr)
N $-$0.3793  0.3732  0.1874,
N  $-$1.4986  0.2410 $-$1.6532, and
Mo  1.5080  0.5375  3.4271), 
while its magnitude was confined to lie between $-$2 and 2 elementary charges. The overall optimization time was several days 
(the parallel version of {\sc Adf} was employed on four Intel Xeon E3-1240 CPU cores). It was found that a point charge of $-0.609325$\,e 
can minimize the overall gradient of the Mo--N$_2$ fragment from  7.51\textperiodcentered10$^{-2}$\,hartree/bohr to 
5.18\textperiodcentered10$^{-3}$\,hartree/bohr. The optimized position of this point charge with respect to the central fragment 
is 488.5\,pm from the distal nitrogen atom away at an N--N--charge angle of 131$^\circ$.

\begin{figure}[H]
 \centering
 \includegraphics[scale=0.05]{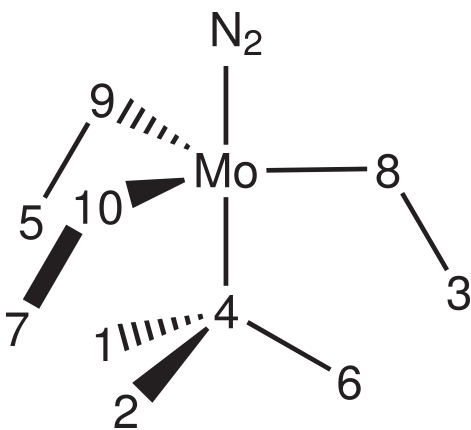}
\caption{Positioning and numbering of the ten point charges (see also Table \ref{tab:pc}).}
 \label{fig:pc}
\end{figure}

Interestingly, the point charge is rather far away from the central fragment, namely 488.5\,pm from the terminal nitrogen atom. 
One could now try to further decrease the gradient by adding a second point charge while keeping the first point charge fixed. 
However, the overall gradient cannot be decreased any further, but instead slightly increases to 5.21\textperiodcentered10$^{-3}$\,hartree/bohr. 
The second point charge is with 618.8\,pm even further away from the distal nitrogen atom than the first one. 

An important issue is the reconstruction of a ligand sphere from a given point charge arrangement. Therefore, in a second exploratory 
study, a total of ten point charges were placed at the positions of the ligand nuclei of a model complex [Mo]$_{\rm O, H}$--N$_2$, in which
the first ligand shell atom is oxygen saturated by a hydrogen atom such that the uncharged [Mo(OH)$_3$NH$_3$N$_2$] complex with Mo in oxidation state +III results
(see Fig.\ \ref{fig:models} below for the structure). 
and kept fixed at these positions. However, during the optimization, a significant number of gradient evaluations did not converge, 
such that no satisfying solution could be obtained. We therefore added the constraint that the total sum of all point charges must 
be zero. With this constraint, we were able to converge a solution within a few days with the standard settings of {\sc Mathematica}. 
The optimal point charges, together with the corresponding overall gradient, is given in Table \ref{tab:pc}. The resulting overall 
gradient on the Mo--N$_2$ fragment is 2.44\textperiodcentered10$^{-2}$\,hartree/bohr. Compared to the previous model system of 
one and two point charges, this is almost one order of magnitude larger. Table \ref{tab:pc} also includes for comparison Hirshfeld 
charges\cite{theoret_chim_acta_1977_44_129} and multipole-derived charges\cite{j_comp_chem_2001_22_79} obtained from a calculation 
on the model complex [Mo]$_{\rm O, H}$--N$_2$. 
For the multipole-derived charges, the atomic multipoles are obtained from the 
electron density up to a given order (in our case quadrupole moments\,---\,the charges are therefore referred to as ``MDC-q'') which 
are then reconstructed exactly by distributing charges at the atom positions.

\begin{table}[H]
\caption{\label{tab:pc} Point charges (in elementary charges) obtained after an optimization with {\sc Mathematica} (second column) 
compared to Hirshfeld and MDC-q charges calculated for complex [Mo]$_{\rm O, H}$--N$_2$ (third and fourth column, respectively). 
The individual positions are shown in Fig.~\ref{fig:pc}. For each set of point charges, the resulting overall gradient is given 
in hartree/bohr in the last line.}
\begin{center}
\begin{tabular}{rrrr}
\hline
position & optimized & Hirshfeld & MDC-q       \\
\hline
1        & $-$0.421  & 0.133     & 0.256       \\
2        & $-$0.542  & 0.125     & 0.261       \\
3        & 0.030     & 0.117     & 0.592       \\
4        & 0.698     & $-$0.169  & $-$0.738    \\
5        & $-$0.029  & 0.117     & 0.603       \\
6        & $-$0.660  & 0.133     & 0.261       \\
7        & 0.339     & 0.095     & 0.514       \\
8        & 0.155     & $-$0.333  & $-$0.883    \\
9        & 0.052     & $-$0.333  & $-$0.894    \\
10       & 0.378     & $-$0.334  & $-$0.818    \\
\hline
gradient & 2.44\textperiodcentered10$^{-2}$ & 8.70\textperiodcentered10$^{-2}$ & 2.00\textperiodcentered10$^{-1}$ \\
\hline
\end{tabular}
\end{center}
\end{table}

We see that the optimized point charges decrease the overall gradient significantly better compared to the Hirshfeld and MDC-q charges. 
Most notably, the overall gradient is even smaller than in the full [Mo]$_{\rm O, H}$--N$_2$ model (see below). The comparison with the 
Hirshfeld and MDC-q charges allows us to make some interesting observations. First of all, we see that the optimized charges 
deviate strongly from both the Hirshfeld as well as the MDC-q charges. Moreover, the symmetry present in the Hirshfeld and MDC-q 
point charge arrangements, originating from the threefold symmetry axis present in the [Mo]$_{\rm O, H}$--N$_2$ model complex, 
can no longer be found in the optimized charges. This would suggest that in each of the three equatorial positions, a different 
ligand would be present. The Hirshfeld and MDC-q charges agree qualitatively with each other in the sense that the more electronegative 
atoms oxygen and nitrogen all have a negative partial charge, with larger negative charges on the oxygen atoms (which has a 
larger electronegativity than nitrogen). All hydrogen atoms have a positive partial charge. Partial charges of the hydrogen 
atoms bound to the oxygen atoms are larger than those in the ammonia ligand, which one would also expect in view 
of the electronegativities of oxygen and nitrogen. If we apply this qualitative reasoning now to the optimized charges, we would 
replace the terminal positions 1, 2, and 6 (positions of the hydrogen atoms of the NH$_3$ ligand) with a strongly electronegative 
element (regarding the valences, fluorine could be a good choice as terminal atom), while at position 5, an atom with a medium 
electronegativity would be placed. Moreover, the possible valences of a given atom type in a given position provide important 
additional information. For example, in position four, an atom from the fifth main group (nitrogen, phosphorus, etc.) would perfectly 
fit into the local coordination environment. From the optimized charge of this position, which is 0.698\,e, we understand that the 
corresponding atom type should have a rather low electronegativity, as only this would be compatible with this charge. One would 
therefore rather employ phosphorus in position four instead of nitrogen. 

Another aspect that we have not touched on so far is the role of the formal oxidation state of a transition metal center within a complex.
In general, the total charge of a complex toegether with the heterolytically cleaved metal--ligand bonds determines the formal
oxidation state. Thus, the total charge of a point-charge field can be considered to determine the formal oxidation state, which
in turn may be used as a constraint in the optimization process. In the case of the Schrock complex, the formal oxidation state is
Mo(III). In a general design process, different oxidation states need to be investigated. At the example of Schrock-type complexes
we had considered oxidation states of +III and 0 for a molybdenum center bound by a tripodal carbene ligand \cite{inorg_chem_2009_48_1638}.
These different oxidation states can be adjusted by pre-defining a fixed total charge of the point-charge field. I.e., if the metal fragment 
is embedded in a point-charge field that has a zero overall charge, then the metal center in the Mo-NN fragment would be considered to be in 
oxidation state 0. For this reason, we always considered the triatomic fragment Mo-NN to be triply positively charged yielding a formal oxidation state
of +III for Mo.
However, we should also stress that the total charge of the point charge field does not necessarily result in a fixation
of the oxidation state at the metal center. Apart from the fact that the oxidation state is a concept with limited validity, we understand the point-charge field as a rather
abstract general environment into which we embed the metal fragment. Hence, the chemical structure that may represent such a point charge field (C-representability)
still needs to be found and only once this has been accomplished, the oxidation state may be assigned. 

Moreover, we should emphasize that the simple 10-point-charge model discussed is only an example of a rather general
discretized electrostatic embedding. It must not be mistaken for an attempt to revive crystal field theory for design purposes. 
Although crystal field theory may be considered a guiding principle, it is constrained to an underlying ligand structure with known
coordination number and structure. Our electrostatic embedding approach is more general in a sense that a point charge of some
value may be put at any position in space, if it reduces the gradient and can be represented by some ligand structure.

\subsection{Direct Optimization by Positioning of Nuclei and Adding Electrons}
\label{sec:adding}

In this final part of the work, we decrease the complexity of the optimization problem by adding meaningful chemical fragments, 
i.e., functional groups, to the central fragment. Still, it is {\it a priori} not known how many electrons and how many nuclei 
(and of which type and at which position) will be required to reduce the gradient. Note that this means that the overall charge 
and spin state remain unknown, and eventually have to be optimized, too. A feasible ansatz is the stepwise build-up of a ligand 
sphere exploiting chemical principles for the scaffold construction. Clearly, the full beauty of inverse design will only flourish 
if this can be done in a first-principles way, but for this first attempt we are advised to exploit chemical knowledge in order 
to identify a feasible strategy.

In a first step, one could start with only a few atoms coordinating directly to the central fragment, thus building up a first 
ligand shell. Due to their close spatial proximity to this central fragment, one can expect that the atoms of this first ligand 
shell will have the largest influence on it. Moreover, a reasonable starting guess for the number and position of these initial 
atoms can be made by exploiting van der Waals or covalent radii and the chemical knowledge about atomic valence. In the present 
case, we can find good starting configurations by studying the coordination numbers and geometries (bond lengths, etc.) of molybdenum 
complexes\cite{wibe07}. In a next step, the position and the type of atoms in the first ligand sphere are optimized such that the 
gradient is minimized considering suitable capping atoms to saturate all valencies. We then proceed by iteratively adding atoms 
to the ones already present, thereby creating an increasingly complex ligand shell, in order to minimize the gradient further. 
Clearly, eventually advanced scaffold construction algorithms will be required to account for all possible chemical situations 
brought about by the increasing ligand shell (e.g. ring closure instead of single extensions and so forth).

\begin{figure}[H]
 \centering
 \includegraphics{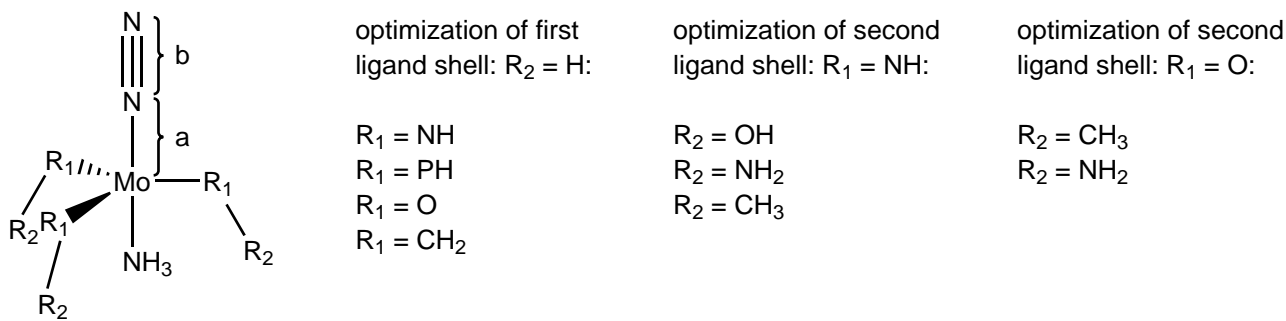}
 \caption{Structural hierarchy of model complexes [Mo]$_{\rm R_1, R_2}$--N$_2$.}
 \label{fig:models}
\end{figure}

We deduced a range of model complexes from the original Schrock complex depicted in Fig.~\ref{fig:models}. All models contain 
the central Mo--N$_2$ fragment; the molecular environment is modelled by two ligand shells with substituents R$_1$ and R$_2$, 
respectively. Accordingly, we denote these models [Mo]$_{\rm R_1, R_2}$--N$_2$, where the subscripts indicate which substituents 
are present in the two ligand shells. For the sake of simplicity, we restrict our model systems to a trigonal-bipyramidal 
coordination geometry as found in the original Schrock catalyst and varied only the type of the three equatorial atoms (their 
position is optimized, too). Any remaining unsaturated valencies are saturated with hydrogen capping atoms.
Note also that the possible groups R$_1$ have been chosen such that the metal center is in oxidation state +III. This restriction
is, however, not mandatory for a general design study.

The choice of the fixed Mo--N and N--N bond lengths (denoted as $a$ and $b$ in Fig.~\ref{fig:models}) is important. There are 
two obvious choices for $a$. Either one chooses a bond length known for similar dinitrogen binding complexes, in this case 
from the original Schrock catalyst (198.4\,pm\cite{inorg_chem_2009_48_1638}), or one can let this bond length relax to its 
equilibrium value in a given model complex. If one sets $b$ to a value that indicates a certain desirable degree of bond 
activation as found in the original Schrock complex (i.e., 114.2\,pm\cite{inorg_chem_2009_48_1638}), the N--N distance 
represents an activated molecular dinitrogen ligand (the N--N distance for isolated N$_2$ is calculated to be 110.4\,pm\cite{inorg_chem_2009_48_1638}). 
If we had intended to find a complex which directly transforms dinitrogen upon binding to a diazenoid or even hydrazinoid species, 
we would set $b$ to a value of 125.2\,pm and 144.9\,pm as calculated for isolated diazene or hydrazine. First, we set both $a$ 
and $b$ to the values they have in the Schrock complex; later in this study, we will investigate different values for these 
two bond lengths. All model complexes have then been optimized in their doublet spin state under the constraint that the 
distances $b$ (and $a$) of Fig.~\ref{fig:models} are kept fixed.

\subsubsection{Constructing a N$_2$-Binding Complex}

For elucidating how to construct a complex which binds molecular dinitrogen, we can set the distances $a$ and $b$ to the values 
found in the original Schrock N$_2$ complex with the HIPT substituents (for the initial positioning of the other atoms, idealized 
bond angles and bond lengths deduced from covalent radii may be used). For this case, Table \ref{tab:grad} lists absolute 
values of the gradients for the model systems mentioned above. According to these data, the overall gradient $|\nabla_{\rm frag}E_{\rm el}|$ 
is decreased significantly in all model systems compared to its value in the isolated central fragment as we would expect 
owing to the saturation of valencies. 
However, while the gradients on the molybdenum atom and on the nitrogen atom attached to 
it are decreased, the gradient on the second nitrogen atom is larger in all model systems compared to the isolated central 
fragment. The model system with phosphorus in the first ligand shell, [Mo]$_{\rm PH,H}$--N$_2$, features larger gradients 
than the other complexes. With a value of 5.64\textperiodcentered10$^{-2}$\,hartree/bohr, it clearly has the largest overall 
gradient of this series of model systems. Apparently, phosphorus does not stabilize the central fragment as well as carbon, 
nitrogen, or oxygen (interestingly, this is in contrast to what we found in section \ref{sec:pointcharges}). 
From these three elements, oxygen is obviously best suited, as [Mo]$_{\rm O,H}$--N$_2$ features the smallest overall gradient (namely, 
3.40\textperiodcentered10$^{-2}$\,hartree/bohr) on the central fragment, although the overall gradient is not much larger if 
nitrogen (as in the original Schrock catalyst) is present instead. The overall gradient on nuclei of the ligand sphere 
(denoted as $|\nabla_{\rm env}E_{\rm el}|$) is very small, since the positions of its atoms are fully optimized (also note 
that since $|\nabla_{\rm env}E_{\rm el}|$ includes all nuclear gradient components of the ligand sphere, it tends to be larger for larger ligand spheres). 
If we would not optimize these positions, and simply use idealized angles and covalent radii to construct a ligand sphere, 
then both, nuclei of the ligand sphere and the central fragment, feature large overall gradients between 10$^{-2}$ and 
10$^{-1}$ hartree/bohr for all model systems. 

We carried out the second-shell construction step with model systems [Mo]$_{\rm O, CH_3}$--N$_2$ and [Mo]$_{\rm O, NH_2}$--N$_2$ 
(see Fig.~\ref{fig:models}) for which we found the lowest absolute gradients. While the overall gradient of 
[Mo]$_{\rm O, NH_2}$--N$_2$ (6.92\textperiodcentered10$^{-2}$\,hartree/bohr) is larger than in its parent model [Mo]$_{\rm O,H}$--N$_2$, 
we can decrease it to 2.16\textperiodcentered10$^{-2}$\,hartree/bohr when binding methyl groups to the oxygen atoms. In a preliminary 
third optimization cycle, we found that a model complex with ethanolate ligands, [Mo]$_{\rm O, CH_2CH_3}$--N$_2$, further decreases 
the overall gradient to 1.57\textperiodcentered10$^{-2}$\,hartree/bohr.

An open question is whether an even smaller final overall gradient could possibly be obtained with a complex which does not feature 
the smallest gradient in the first construction steps. In order to address this question, we investigate model complexes 
[Mo]$_{\rm NH,OH}$--N$_2$, [Mo]$_{\rm NH,NH_2}$--N$_2$, and [Mo]$_{\rm NH,CH_3}$--N$_2$ (cf., Fig.~\ref{fig:models}). The 
parent complex of these systems is [Mo]$_{\rm NH, H}$--N$_2$, which features nitrogen atoms coordinating to the molybdenum atom. 
With a value of approximately 2.83\textperiodcentered10$^{-3}$\,hartree/bohr, [Mo]$_{\rm NH,OH}$--N$_2$ exhibits the smallest 
overall gradient within these three systems (Table \ref{tab:grad}). However, this value is larger than the overall gradient of 
[Mo]$_{\rm O,CH_3}$--N$_2$. Thus, in this case it is not advantageous to start from a non-optimal parent complex in the second 
construction cycle. However, we will present in the next paragraph a different example where in fact the opposite is the case.

\begin{table}[H]
\caption{\label{tab:grad} {\bf N$_2$ binding:} Absolute values of the Cartesian gradients on nuclei of the central fragment (N(1) 
is the nitrogen atom bound to the molybdenum atom) for the different model systems shown in Fig.~\ref{fig:models}. In all these 
model systems, the distances $a$ and $b$ (cf., Fig.~\ref{fig:models}) have been fixed to the values found in the full Schrock 
complex depicted in Fig.~\ref{fig:schrock}. All data is given in hartree/bohr.}
\begin{center}
\begin{tabular}{lrrrrr}
\hline
model & $|\nabla_{\rm Mo}E_{\rm el}|$ & $|\nabla_{\rm N(1)}E_{\rm el}|$ & $|\nabla_{\rm N(2)}E_{\rm el}|$ & $|\nabla_{\rm frag}E_{\rm el}|$ & $|\nabla_{\rm env}E_{\rm el}|$ \\
\hline
Mo--N$_2$ & 3.76\textperiodcentered10$^{-2}$ & 3.48\textperiodcentered10$^{-2}$ & 2.73\textperiodcentered10$^{-3}$ & 7.51\textperiodcentered10$^{-2}$ & 0.00 \\
full Schrock &  3.56\textperiodcentered10$^{-3}$ & 2.32\textperiodcentered10$^{-3}$ & 7.06\textperiodcentered10$^{-3}$ & 1.17\textperiodcentered10$^{-2}$ & 8.74\textperiodcentered10$^{-2}$\\
\hline
{[}Mo]$_{\rm O,H}$--N$_2$ & 1.40\textperiodcentered10$^{-3}$ & 1.68\textperiodcentered10$^{-2}$ & 1.58\textperiodcentered10$^{-2}$ & {\bf 3.40\textperiodcentered10$^{-2}$} & 2.42\textperiodcentered10$^{-3}$\\
{[}Mo]$_{\rm NH,H}$--N$_2$ & 8.47\textperiodcentered10$^{-3}$ & 2.11\textperiodcentered10$^{-2}$ & 1.26\textperiodcentered10$^{-2}$ & 3.45\textperiodcentered10$^{-2}$ & 9.28\textperiodcentered10$^{-3}$\\
{[}Mo]$_{\rm CH_2,H}$--N$_2$ & 5.37\textperiodcentered10$^{-3}$ & 2.04\textperiodcentered10$^{-2}$ & 1.50\textperiodcentered10$^{-2}$ & 4.18\textperiodcentered10$^{-2}$ & 4.19\textperiodcentered10$^{-3}$\\
{[}Mo]$_{\rm PH,H}$--N$_2$ & 1.10\textperiodcentered10$^{-2}$ & 2.81\textperiodcentered10$^{-2}$ & 1.73\textperiodcentered10$^{-2}$ & 5.64\textperiodcentered10$^{-2}$ & 3.31\textperiodcentered10$^{-3}$\\
\hline
{[}Mo]$_{\rm O,CH_3}$--N$_2$ & 1.86\textperiodcentered10$^{-3}$ & 9.40\textperiodcentered10$^{-3}$ & 1.03\textperiodcentered10$^{-2}$ & {\bf 2.16\textperiodcentered10$^{-2}$} & 1.05\textperiodcentered10$^{-2}$\\
{[}Mo]$_{\rm O,NH_2}$--N$_2$ & 2.16\textperiodcentered10$^{-2}$ & 3.45\textperiodcentered10$^{-2}$ & 1.31\textperiodcentered10$^{-2}$ & 6.92\textperiodcentered10$^{-2}$ & 6.18\textperiodcentered10$^{-3}$\\
\hline
{[}Mo]$_{\rm NH,OH}$--N$_2$ & 1.39\textperiodcentered10$^{-2}$ & 1.35\textperiodcentered10$^{-2}$ & 9.23\textperiodcentered10$^{-4}$ & {\bf 2.83\textperiodcentered10$^{-2}$} & 6.46\textperiodcentered10$^{-3}$\\
{[}Mo]$_{\rm NH,NH_2}$--N$_2$ & 1.21\textperiodcentered10$^{-3}$ & 1.56\textperiodcentered10$^{-2}$ & 1.72\textperiodcentered10$^{-2}$ & 3.40\textperiodcentered10$^{-2}$ & 5.84\textperiodcentered10$^{-3}$\\
{[}Mo]$_{\rm NH,CH_3}$--N$_2$ & 1.25\textperiodcentered10$^{-2}$ & 3.47\textperiodcentered10$^{-2}$ & 2.22\textperiodcentered10$^{-2}$ & 6.94\textperiodcentered10$^{-2}$ & 5.06\textperiodcentered10$^{-3}$\\
\hline
\end{tabular}
\end{center}
\end{table}

In order to investigate the flexibility and capabilities of the shell-wise construction approach, we let the Mo--N distance $a$ 
(see Fig.~\ref{fig:models}) relax during the optimization procedure instead of keeping it fixed. With this additional degree 
of freedom, we can expect to find smaller overall gradients compared to the cases studied above where $a$ was kept fixed. 
The resulting data are given for all model systems in Table \ref{tab:grad2}. We observe the same general trend as before. 
Oxygen atoms in the first ligand shell lead to the smallest overall gradient, although nitrogen atoms are also about equally 
good, while phosphorus is again not well suited to decrease the gradient. In the second ligand shell, the methyl groups decrease 
the overall gradient further. Nevertheless, we note that now [Mo]$_{\rm NH,OH}$--N$_2$ features with 1.42\textperiodcentered10$^{-2}$\,hartree/bohr 
clearly a smaller overall gradient than [Mo]$_{\rm O,CH_3}$--N$_2$ with 2.01\textperiodcentered10$^{-2}$\,hartree/bohr. This 
shows that it is indeed possible to obtain a smaller overall gradient by starting from a parent complex which does not feature 
the smallest gradient.

However, we also note that for the [Mo]$_{\rm NH, OH}$--N$_2$ model system, the Mo--N distance $a$ is quite long with 201.1\,pm. 
The [Mo]$_{\rm OH,CH_3}$--N$_2$ model system, on the contrary, features a Mo--N bond length which is very similar to the one 
found in the full Schrock complex. We calculated an intrinsic binding energy, from the structures of both the dinitrogen ligand 
as well as the model complexes without the dinitrogen ligand kept as in the complex (E$_{\rm bind,\,N_2}^{\rm (fixed)}$), and the (true) binding 
energy from both ``reactants'' which were structurally fully optimized (E$_{\rm bind,\,N_2}^{\rm (relaxed)}$). In addition, we calculated the binding 
energy for the case where only N$_2$ was allowed o relax (E$_{\rm bind,\,N_2}^{\rm (N_2~relaxed)}$). Not unexpected, in the 
calculation of E$_{\rm bind,\,N_2}^{\rm (relaxed)}$ many of the fragments experienced significant structural changes during 
relaxation. [Mo]$_{\rm O,NH_2}$ and [Mo]$_{\rm NH,OH}$ even decomposed completely. Therefore, we will omit these binding energies 
in the following dicussion and rather focus on the intrinsic binding energies. We see that the values for E$_{\rm bind,\,N_2}^{\rm (fixed)}$ 
is always larger than the corresponding values for E$_{\rm bind,\,N_2}^{\rm (N_2~relaxed)}$, but the qualitative trends are the 
same in both cases.

\begin{table}[H]
\caption{\label{tab:grad2} {\bf N$_2$ binding:} Absolute values of the Cartesian gradients on nuclei of the central fragment 
(N(1) is the nitrogen atom bound to the molybdenum atom) and Mo--N distance $a$ for the different model systems shown in Fig.~\ref{fig:models}. 
In all these model systems, the distance $b$ (cf., Fig.~\ref{fig:models}) has been fixed to the value found in the full Schrock complex 
depicted in Fig.~\ref{fig:schrock}, while $a$ was allowed to relax. All data is given in hartree/bohr, if not stated otherwise.}
\begin{center}
\begin{tabular}{lrrrrrr}
\hline
model & $|\nabla_{\rm Mo}E_{\rm el}|$ & $|\nabla_{\rm N(1)}E_{\rm el}|$ & $|\nabla_{\rm N(2)}E_{\rm el}|$ & $|\nabla_{\rm frag}E_{\rm el}|$ & $|\nabla_{\rm env}E_{\rm el}|$ & $a$ / pm \\
\hline
Mo--N$_2$ & 2.5\textperiodcentered10$^{-5}$ & 2.21\textperiodcentered10$^{-2}$ & 2.22\textperiodcentered10$^{-2}$ & 4.44\textperiodcentered10$^{-2}$ & 0.00 & 206.9 \\
full Schrock &  1.04\textperiodcentered10$^{-4}$ & 9.47\textperiodcentered10$^{-4}$ & 2.06\textperiodcentered10$^{-4}$ & 1.26\textperiodcentered10$^{-3}$ & 3.80\textperiodcentered10$^{-2}$ & 199.0  \\
\hline
{[}Mo]$_{\rm O,H}$--N$_2$ & 5.70\textperiodcentered10$^{-4}$ & 1.64\textperiodcentered10$^{-2}$ & 1.63\textperiodcentered10$^{-2}$ & {\bf 3.33\textperiodcentered10$^{-2}$} & 1.51\textperiodcentered10$^{-3}$ & 198.6 \\
{[}Mo]$_{\rm NH,H}$--N$_2$ & 8.43\textperiodcentered10$^{-4}$ & 1.68\textperiodcentered10$^{-2}$ & 1.67\textperiodcentered10$^{-2}$ & 3.44\textperiodcentered10$^{-2}$ & 4.39\textperiodcentered10$^{-3}$ & 196.9 \\
{[}Mo]$_{\rm CH_2,H}$--N$_2$ & 9.69\textperiodcentered10$^{-4}$ & 1.79\textperiodcentered10$^{-2}$ & 1.77\textperiodcentered10$^{-2}$ & 3.66\textperiodcentered10$^{-2}$ & 2.93\textperiodcentered10$^{-3}$ & 197.4 \\
{[}Mo]$_{\rm PH,H}$--N$_2$ & 9.19\textperiodcentered10$^{-4}$ & 2.39\textperiodcentered10$^{-2}$ & 2.38\textperiodcentered10$^{-2}$ & 4.86\textperiodcentered10$^{-2}$ & 8.24\textperiodcentered10$^{-3}$ & 200.4 \\
\hline
{[}Mo]$_{\rm O,CH_3}$--N$_2$ & 9.91\textperiodcentered10$^{-4}$ & 1.01\textperiodcentered10$^{-2}$ & 9.00\textperiodcentered10$^{-3}$ & {\bf 2.01\textperiodcentered10$^{-2}$} & 1.05\textperiodcentered10$^{-2}$& 198.0 \\
{[}Mo]$_{\rm O,NH_2}$--N$_2$ & 1.25\textperiodcentered10$^{-3}$ & 2.36\textperiodcentered10$^{-2}$ & 2.41\textperiodcentered10$^{-2}$ & 4.90\textperiodcentered10$^{-2}$ & 8.85\textperiodcentered10$^{-3}$ & 202.9 \\
\hline
{[}Mo]$_{\rm NH,OH}$--N$_2$ & 1.15\textperiodcentered10$^{-3}$ & 6.58\textperiodcentered10$^{-3}$ & 6.51\textperiodcentered10$^{-3}$ & {\bf 1.42\textperiodcentered10$^{-2}$} & 7.02\textperiodcentered10$^{-3}$ & 201.1\\
{[}Mo]$_{\rm NH,NH_2}$--N$_2$ & 1.29\textperiodcentered10$^{-3}$ & 1.61\textperiodcentered10$^{-2}$ & 1.70\textperiodcentered10$^{-2}$ & 3.44\textperiodcentered10$^{-2}$ & 9.07\textperiodcentered10$^{-3}$ & 198.5 \\
{[}Mo]$_{\rm NH,CH_3}$--N$_2$ & 1.07\textperiodcentered10$^{-3}$ & 2.84\textperiodcentered10$^{-2}$ & 2.84\textperiodcentered10$^{-2}$ & 5.79\textperiodcentered10$^{-2}$ & 5.69\textperiodcentered10$^{-3}$ & 196.1\\
\hline
\end{tabular}
\end{center}
\end{table}

\begin{table}[H]
\caption{\label{tab:energies} {\bf N$_2$ binding:} N$_2$ and ligand (i.e., the ligand in the equatorial position) binding energies 
for the different model systems shown in Fig.~\ref{fig:models}. For the intrinsic binding energies E$_{\rm bind,\,N_2}^{\rm (fixed)}$ 
and E$_{\rm bind,\,ligand}^{\rm (fixed)}$, the (dinitrogen) ligand as well as the complex fragment were kept fixed, while they were 
fully relaxed in the calculation of E$_{\rm bind,\,N_2}^{\rm (relaxed)}$. Finally, in the case of E$_{\rm bind,\,N_2}^{\rm (N_2~relaxed)}$ 
only dinitrogen was optimized, while the model system fragment was kept fixed. In all these model systems, the distance $b$ 
(cf., Fig.~\ref{fig:models}) has been fixed to the value found in the full Schrock complex depicted in Fig.~\ref{fig:schrock}, while 
$a$ was allowed to relax (with the exception of E'$_{\rm bind,\,ligand}^{\rm (fixed)}$, where also $a$ was kept fixed). All data 
are given in kJ\,mol$^{-1}$.}
\begin{center}
\begin{tabular}{lrrrrr}
\hline
model                          & E$_{\rm bind,\,N_2}^{\rm (fixed)}$ & E$_{\rm bind,\,N_2}^{\rm (relaxed)}$ & E$_{\rm bind,\,N_2}^{\rm (N_2~relaxed)}$ & E$_{\rm bind,\,ligand}^{\rm (fixed)}$ & E'$_{\rm bind,\,ligand}^{\rm (fixed)}$\\
\hline
full Schrock                   & n/a                           & $-$151.8$^{a}$              & n/a     & n/a      & n/a \\
\hline
{[}Mo]$_{\rm O,H}$--N$_2$     & $-$155.9                      & $-$119.5                    & $-$147.5   & $-$941.6    & $-$941.7 \\
{[}Mo]$_{\rm NH,H}$--N$_2$    & $-$187.3                      & $-$137.7                    & $-$178.9   & $-$914.2    & $-$914.0 \\
{[}Mo]$_{\rm CH_2,H}$--N$_2$  & $-$183.2                      & $-$146.6                    & $-$174.5   & $-$912.6    & $-$911.2 \\
{[}Mo]$_{\rm PH,H}$--N$_2$    & $-$150.4                      & $-$117.0                    & $-$141.9   & $-$789.8    & $-$791.3 \\
\hline
{[}Mo]$_{\rm O,CH_3}$--N$_2$  & $-$158.0                      & $-$126.1                    & $-$149.5   & $-$859.8    & $-$860.1 \\
{[}Mo]$_{\rm O,NH_2}$--N$_2$  & $-$134.6                      & 237.0$^{b}$                 & $-$126.1   & $-$919.5    & $-$921.5 \\
\hline
{[}Mo]$_{\rm NH,OH}$--N$_2$   & $-$154.4                      & 71.0$^{b}$                  & $-$146.0   & $-$941.6    & $-$942.3 \\
{[}Mo]$_{\rm NH,NH_2}$--N$_2$ & $-$177.5                      & $-$129.9                    & $-$169.1   & $-$919.5     & $-$919.5 \\
{[}Mo]$_{\rm NH,CH_3}$--N$_2$ & $-$194.7                      & $-$136.1                    & $-$186.3   & $-$863.8     & $-$862.1 \\
\hline
\end{tabular}
\end{center}
$^{a}$ This value was obtained from Ref.~\cite{inorg_chem_2009_48_1638} and was calculated with a slightly different 
methodology (see Ref.~\cite{inorg_chem_2009_48_1638}). However, since in both methodologies a high degree of accuracy 
was aimed at, the differences of the actual numerical results are negligible (for example, the distance $a$ was calculated 
to be 198.4\,pm in Ref.~\cite{inorg_chem_2009_48_1638}, while it is found to be 199.0\,pm in this work).

$^{b}$ In these two cases, the model system fragment decomposes during relaxation.
\end{table}

When comparing the binding energies E$_{\rm bind,\,N_2}^{\rm (fixed)}$ of the two systems [Mo]$_{\rm NH,OH}$--N$_2$ and 
[Mo]$_{\rm OH,CH_3}$--N$_2$ (see Table \ref{tab:grad2}), we understand that both values are slightly larger than the binding 
energy observed in the full Schrock complex. The binding energy of the [Mo]$_{\rm NH,OH}$--N$_2$ model complex 
(154.4\,kJ\,mol$^{-1}$) is even more similar to the one of the Schrock catalyst (at least regarding the true intrinsic binding 
energies\,---\,this picture changes when considering E$_{\rm bind,\,N_2}^{\rm (N_2~relaxed)}$). We therefore conclude that 
also the [Mo]$_{\rm NH,OH}$--N$_2$ complex activates the N$_2$ fragment to a sufficient amount even though the Mo--N distance 
$a$ of 201.1\,pm might suggest otherwise. Therefore, we see that it is possible to find a smaller overall gradient by starting 
from a complex which was not optimal in the previous optimization cycle. This implies that no primary parts of a chelate-ligand 
scaffold may be disregarded in early optimization steps. Hence, the computational effort should not be reduced by eliminating 
intermediary scaffolds too early.

We calculated also the intrinsic binding energies of the ligands in the equatorial position in order to understand whether the 
gradient-reducing effect of oxygen atoms in the first ligand shell is a viable target for synthetic attempts; these data are 
also given in Table \ref{tab:energies}. The reason for the large values of these binding energies is the fact that these ligands 
are negatively charged, which leads to a separation of positive (on molybdenum) and negative charges in this artificial dissociation 
process of isolated species. The hydroxide ligand of the [Mo]$_{\rm O,H}$--N$_2$ complex, which leads to the smallest overall 
gradient, is most strongly bound to molybdenum (941.6\,kJ\,mol$^{-1}$), while it is much smaller (789.8\,kJ\,mol$^{-1}$) for 
[Mo]$_{\rm PH,H}$--N$_2$. The amido ligand in the model system [Mo]$_{\rm NH,H}$--N$_2$ is bound with 914.0\,kJ\,mol$^{-1}$. 
However, in the second optimization cycle, this relation is not strictly valid anymore. The methanolate ligand in [Mo]$_{\rm O,CH_3}$--N$_2$ 
clearly decreases the gradient compared to the parent system [Mo]$_{\rm O,H}$--N$_2$, but it is less strongly bound (859.8\,kJ\,mol$^{-1}$). 
On the other hand, the ONH$_2$ ligand (considered as a monoanion), which leads to a very large overall gradient, is rather strongly bound with 919.5\,kJ\,mol$^{-1}$.

\begin{figure}[H]
\linespread{0.66}
 \centering
 \includegraphics{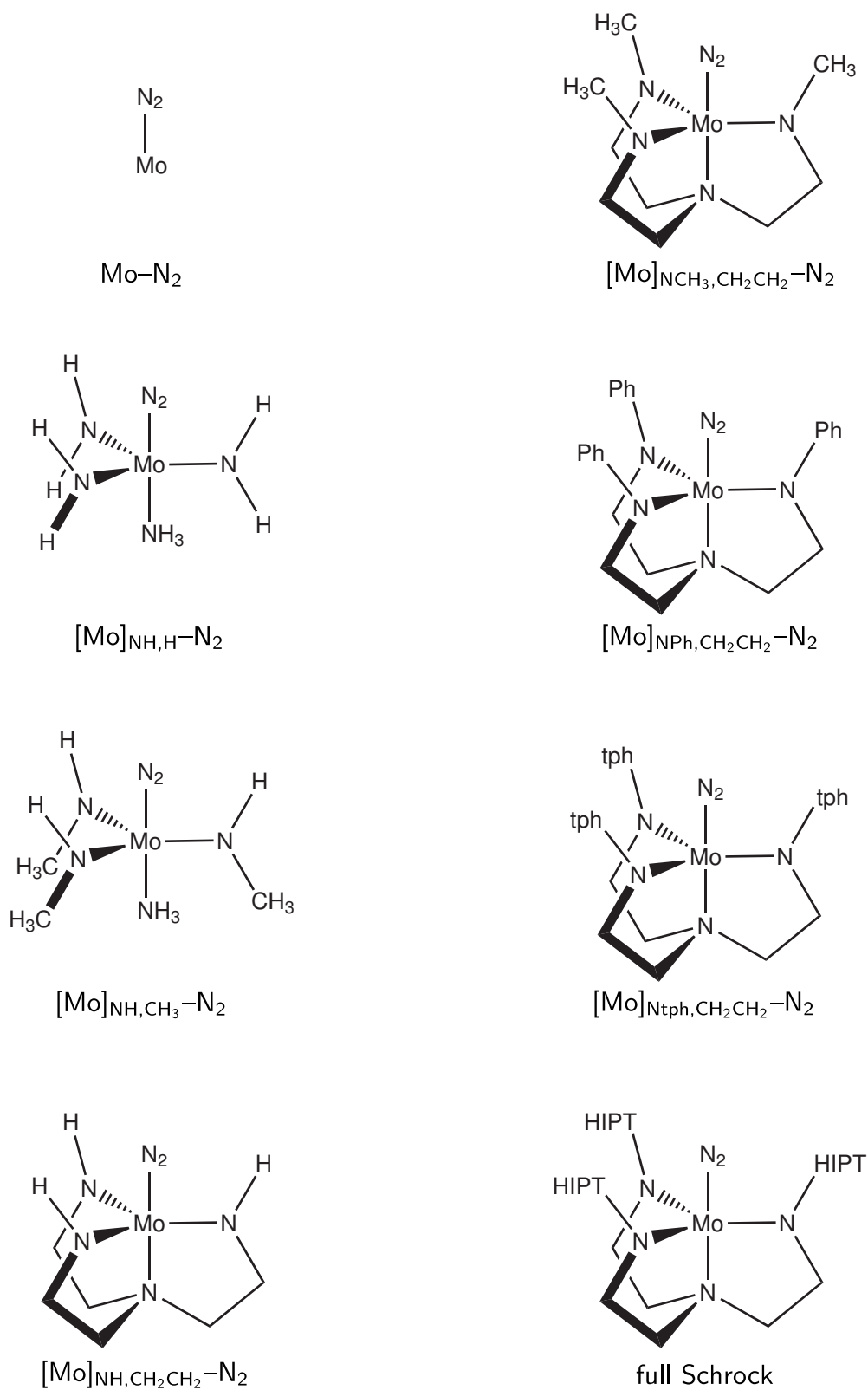}
 \caption{Overview of the model system series gradually approximating the full Schrock complex depicted in the bottom right corner. 
The abbreviation ``tph'' stands for terphenyl (i.e., the HIPT ligand without the isopropyl substituents).}
 \label{fig:schrock_series}
\end{figure}

In order to further investigate how many optimization cycles might be necessary to decrease the gradient on the central fragment to 
an acceptable threshold, we constructed a series of model systems gradually approaching the full Schrock complex as depicted in 
Fig.~\ref{fig:schrock_series}. We then optimized the positions of all atoms except the ones of the central fragment so that both 
distances $a$ and $b$ were kept fixed. The resulting gradients on this central fragment are given in Table \ref{tab:grad_schrock}. 
One can see that the overall gradient does not monotonically decrease when enlarging the model systems. On the contrary, it even 
reaches a value which is higher than the one obtained in the original Mo--N$_2$ fragment. A similar behavior is observed for the 
gradients on the individual atoms. The largest two model systems feature overall gradients which are only slightly larger than the 
one of the full Schrock complex. We also note that the overall gradient of the full Schrock complex drops by one order of magnitude 
if the Mo-N and N--N distances are allowed to relax. Upon relaxation, the Mo--N distance is elongated from 198.4 to 199.0\,pm, 
while the N--N distance remains almost unchanged. The value of 198.4\,pm for the Mo--N distance was taken from Ref.~\cite{inorg_chem_2009_48_1638}, 
where a different basis set (namely, Ahlrichs' GTO-type TZVP basis\cite{j_chem_phys_1994_100_5829} at all atoms except for carbon and 
hydrogen, for which the SVP basis\cite{j_chem_phys_1992_97_2571} was chosen). Since in our calculations the Slater-type TZP basis set 
is employed, the equilibrium distance of the Mo--N distance is slightly different (by 0.6\,pm). Using a Mo--N bond length of 198.4\,pm 
therefore leads to overall gradients which are rather large at first sight. However, this observation is important as such a bond 
length would have to be guessed for a yet unknown complex to be constructed.

\begin{table}[H]
\caption{\label{tab:grad_schrock} {\bf N$_2$ binding:} Absolute values of the Cartesian gradients on nuclei of the central 
fragment (last three columns; N(1) is the nitrogen atom bound to the molybdenum atom) for the model systems gradually approaching 
the full Schrock complex (see Fig.~\ref{fig:schrock_series}). In all model systems, the Mo--N distance $a$ is constrained to 198.4\,pm,
 while the N--N distance $b$ is kept fixed at 114.2\,pm. For the full Schrock catalyst, data for both constrained as well as unconstrained 
distances $a$ and $b$ is given. All data is given in hartree/bohr.}
\begin{center}
\begin{tabular}{lrrrrr}
\hline
model                   & $|\nabla_{\rm Mo}E_{\rm el}|$ & $|\nabla_{\rm N(1)}E_{\rm el}|$ & $|\nabla_{\rm N(2)}E_{\rm el}|$ & $|\nabla_{\rm frag}E_{\rm el}|$ & $|\nabla_{\rm env}E_{\rm el}|$ \\
\hline
Mo--N$_2$ & 3.76\textperiodcentered10$^{-2}$ & 3.48\textperiodcentered10$^{-2}$ & 2.73\textperiodcentered10$^{-3}$ & 7.51\textperiodcentered10$^{-2}$ & 0.00\\
{[}Mo]$_{\rm NH,H}$--N$_2$       & 8.47\textperiodcentered10$^{-3}$  & 2.11\textperiodcentered10$^{-2}$ & 1.26\textperiodcentered10$^{-2}$ & 3.45\textperiodcentered10$^{-2}$ & 9.28\textperiodcentered10$^{-3}$\\
{[}Mo]$_{\rm NH,CH_3}$--N$_2$    & 1.30\textperiodcentered10$^{-2}$  & 3.39\textperiodcentered10$^{-2}$ & 2.14\textperiodcentered10$^{-2}$ & 6.82\textperiodcentered10$^{-2}$ & 5.06\textperiodcentered10$^{-3}$\\
{[}Mo]$_{\rm NH,CH_2CH_2}$--N$_2$& 2.74\textperiodcentered10$^{-3}$  & 3.25\textperiodcentered10$^{-2}$ & 2.97\textperiodcentered10$^{-2}$ & 6.49\textperiodcentered10$^{-2}$ & 4.31\textperiodcentered10$^{-3}$\\
{[}Mo]$_{\rm NCH_3,CH_2CH_2}$--N$_2$ & 7.65\textperiodcentered10$^{-4}$  & 3.84\textperiodcentered10$^{-2}$ & 3.79\textperiodcentered10$^{-2}$ & 7.70\textperiodcentered10$^{-2}$ & 9.04\textperiodcentered10$^{-3}$\\
{[}Mo]$_{\rm NPh,CH_2CH_2}$--N$_2$   & 7.37\textperiodcentered10$^{-3}$ & 3.17\textperiodcentered10$^{-3}$ & 4.14\textperiodcentered10$^{-3}$ & 1.47\textperiodcentered10$^{-2}$ & 7.29\textperiodcentered10$^{-3}$\\
{[}Mo]$_{\rm Ntph,CH_2CH_2}$--N$_2$  & 7.46\textperiodcentered10$^{-3}$ & 1.67\textperiodcentered10$^{-3}$ & 5.98\textperiodcentered10$^{-3}$ & 1.51\textperiodcentered10$^{-2}$ & 2.44\textperiodcentered10$^{-2}$\\
full Schrock & 3.56\textperiodcentered10$^{-3}$ & 2.32\textperiodcentered10$^{-3}$ & 7.06\textperiodcentered10$^{-3}$ & 1.17\textperiodcentered10$^{-2}$ & 8.74\textperiodcentered10$^{-2}$\\
full Schrock ($a$, $b$ relaxed) & 1.04\textperiodcentered10$^{-4}$ & 9.47\textperiodcentered10$^{-4}$ & 2.06\textperiodcentered10$^{-4}$ & 1.26\textperiodcentered10$^{-3}$ & 3.80\textperiodcentered10$^{-2}$\\
\hline
\end{tabular}
\end{center}
\end{table}

\subsubsection{Constructing a N$_2$-Activating Complex}

The N--N bond length $b$ (cf., Fig.~\ref{fig:models}) is a convenient descriptor to (pre)define the degree of activation of 
the dinitrogen molecule. In order to find a complex which does not only bind, but also activate molecular nitrogen, one can 
carry out a similar optimization procedure as laid out above, setting $b$ to a value as found in diazene or hydrazine. For 
isolated diazene $b$ we obtained 125.2\,pm while it has a value of 144.9\,pm for isolated hydrazine. Here, we study gradients 
of the models [Mo]$_{\rm NH,H}$--N$_2$, [Mo]$_{\rm PH,H}$--N$_2$, [Mo]$_{\rm O,H}$--N$_2$, and [Mo]$_{\rm CH_3,H}$--N$_2$, 
setting $b$ arbitrarily to a value of 130\,pm. This corresponds to a significant activation; very similar values are found for 
the N--N bond length in the original Schrock complex and derivatives thereof when dinitrogen has been doubly protonated and 
reduced\cite{inorg_chem_2009_48_1638}. The resulting gradient data are given in Table \ref{tab:grad3}. 

\begin{table}[H]
\caption{\label{tab:grad3} {\bf N$_2$ activation:} Absolute values of the Cartesian gradients on nuclei of the central fragment 
(last three columns; N(1) is the nitrogen atom bound to the molybdenum atom) for the different model systems studied. In all these 
model systems, the distance $a$ (cf., Fig.~\ref{fig:models} has been fixed to the value found in the Schrock complex, while $b$ 
was set to 130\,pm. All data are given in hartree/bohr.}
\begin{center}
\begin{tabular}{lrrrrr}
\hline
model & $|\nabla_{\rm Mo}E_{\rm el}|$ & $|\nabla_{\rm N(1)}E_{\rm el}|$ & $|\nabla_{\rm N(2)}E_{\rm el}|$ & $|\nabla_{\rm frag}E_{\rm el}|$ & $|\nabla_{\rm env}E_{\rm el}|$ \\
\hline
Mo--N$_2$ & 8.43\textperiodcentered10$^{-3}$ & 3.11\textperiodcentered10$^{-1}$ & 3.19\textperiodcentered10$^{-1}$ & 6.39\textperiodcentered10$^{-1}$  & 0.00 \\
\hline
{[}Mo]$_{\rm NH,H}$--N$_2$ & 5.11\textperiodcentered10$^{-2}$ & 2.77\textperiodcentered10$^{-1}$ & 3.28\textperiodcentered10$^{-1}$ & 6.56\textperiodcentered10$^{-1}$ & 4.56\textperiodcentered10$^{-3}$ \\
{[}Mo]$_{\rm PH,H}$--N$_2$ & 3.34\textperiodcentered10$^{-2}$ & 3.19\textperiodcentered10$^{-1}$ & 3.52\textperiodcentered10$^{-1}$ & 7.05\textperiodcentered10$^{-1}$ & 2.92\textperiodcentered10$^{-3}$ \\
{[}Mo]$_{\rm O,H}$--N$_2$ & 4.66\textperiodcentered10$^{-2}$ & 3.06\textperiodcentered10$^{-1}$ & 3.52\textperiodcentered10$^{-1}$ & 7.04\textperiodcentered10$^{-1}$  & 4.31\textperiodcentered10$^{-3}$\\
{[}Mo]$_{\rm CH_2,H}$--N$_2$ & 4.91\textperiodcentered10$^{-2}$ & 2.75\textperiodcentered10$^{-1}$ & 3.24\textperiodcentered10$^{-1}$ & 6.49\textperiodcentered10$^{-1}$ & 7.74\textperiodcentered10$^{-3}$\\
\hline
{[}Mo]$_{\rm NH,H}$--N$_2{}^-$ & 6.74\textperiodcentered10$^{-2}$ & 2.08\textperiodcentered10$^{-1}$ & 2.76\textperiodcentered10$^{-1}$ & 5.51\textperiodcentered10$^{-1}$ & 2.73\textperiodcentered10$^{-3}$\\
{[}Mo]$_{\rm NH,H}$--N$_2$--H$^+$ & 1.02\textperiodcentered10$^{-1}$ & 8.58\textperiodcentered10$^{-2}$ & 1.79\textperiodcentered10$^{-1}$ & 3.67\textperiodcentered10$^{-1}$ & 6.41\textperiodcentered10$^{-3}$\\
{[}Mo]$_{\rm NH,H}$--N$_2$--H & 1.51\textperiodcentered10$^{-1}$ & 4.28\textperiodcentered10$^{-2}$ & 1.19\textperiodcentered10$^{-1}$ & 3.13\textperiodcentered10$^{-1}$ & 3.38\textperiodcentered10$^{-3}$\\
\hline
\end{tabular}
\end{center}
\end{table}

As can be seen, our optimization procedure does not have any success: only the absolute gradient on the nitrogen atom bound to 
the molybdenum center can be decreased somewhat, but the resulting overall gradients are all slightly larger (0.1\textperiodcentered10$^{-1}$\,--\,0.5\textperiodcentered10$^{-1}$\,hartree/bohr) than in the free Mo--N$_2$ fragment. However, 
this result is exactly what one should have expected as our model systems are inspired by the original Schrock complex, which itself 
does not activate dinitrogen to such a large extent. The activation of nitrogen is achieved by the Schrock catalyst in a sequence of 
protonation and reduction steps. It was shown for iron fragments clamping an N$_2$ ligand that population of the antibonding orbitals 
on the N$_2$ ligand can activate the bond and induce a structural rearrangement to a diazenoid species\cite{chem_eur_j_2004_10_4443,
chem_eur_j_2005_11_574}. In fact, we find that the overall gradient does already slightly decrease if we reduce the model complex 
[Mo]$_{\rm NH,H}$--N$_2$ by one elementary charge, thus forming [Mo]$_{\rm NH,H}$--N$_2{}^-$. In the original Schrock catalyst, 
the N$_2$ fragment is subject to a significant distortion upon activation, which we have not taken into account so far, i.e., the 
diazenoid and hydrazenoid structures have Mo--N--N angles of about 120$^\circ$. We can thus modify our model systems accordingly 
in order to decrease the overall gradient. However, we found that the [Mo]$_{\rm NH,H}$--N$_2$ complex does not support an Mo--N--N 
angle of 120$^\circ$. Also in the [Mo]$_{\rm NH,H}$--N$_2{}^{-}$ and in the [Mo]$_{\rm NH,H}$--N$_2{}^{2-}$ complex the Mo--N--N 
angle is stable only at 180$^\circ$. Therefore, in this specific case, this structural modification does not lead to a lower overall 
gradient. However, if we protonate the complex at the terminal nitrogen atom of the N$_2$ fragment instead of reducing it, the 
overall gradient is significantly reduced. The gradient can be further optimized by employing a combined reduction and protonation; 
the overall absolute gradient of [Mo]$ _{\rm NH,H}$--N$_2$--H is only about half as large as it is in the original fragment. 
Therefore, these preliminary calculations show that also in this more challenging case, our method can be applied to decrease the 
gradients on all nuclei.

\section{Conclusion and Outlook}

In this work, we have elaborated on a new idea for rational compound design, namely Gradient-driven Molecule Construction (GdMC). 
This approach has been applied to the case of a small-molecule activating catalyst. We started with a predefined central fragment 
and searched for a ligand sphere stabilizing this fragment, i.e., nullifying the nuclear gradient on the fragment (and also on the 
chemical environment) such that this conformation is stable.

We have investigated different approaches for finding such a ligand sphere. We have seen that the overall gradient on the Mo--N$_2$ 
fragment can be signficantly decreased already with only a single point charge. Furthermore, for more complicated point charge 
arrangements, the concept of electronegativity can be used to deduce atom types from a given charge magnitude. Even though an 
assignment of atom types clearly needs to be further elaborated on, the greater challenge is most likely the optimization algorithm 
itself. A local optimization depends strongly on the starting conditions, which means that the quality of the inital guess is decisive 
for the outcome of such an optimization. A global optimization is extremely time-consuming already for very small search spaces. 
The global optimization of a large set of point charges can thus be expected not to be possible in a straightforward fashion. When 
several point charges are utilized, one can use the concept of electronegativity for the assignment of an atomic nucleus to a 
given point charge. One could in principle also imagine a brute-force approach which employs a large number of point charges, such 
that the ligand sphere can be represented as a discretized charge distribution. If this discretized charge distribution can be 
split into an electronic and a nuclear component (obeying the physical fact that the nuclear one will be discrete, while 
the electronic one may extend in all space), one could reconstruct a ligand sphere by analyzing the cusps. For feasibility 
reasons, such an ansatz should exploit analytical knowledge --- such as the asymptotically exponential decay of the electronic charge 
distribution --- in the optimization procedure to an utmost extent.

We have also directly supplied nuclei and electrons to the environment of a central fragment and optimized their number and, in the case of the nuclei, 
their type and spatial location. The resulting optimization problem is highly nontrivial. Still, our first attempts turned out to be promising. The 
most important next step would be to automatize this method such that it can be combined with an established optimization method (such 
as the differential evolution algorithm employed here), which would allow us to explore the limits of this approach. In this 
respect, the automatic generation of reasonable starting structures appears imperative. Such a structure generation could rely on 
standard covalent radii and idealized bond and dihedral angles in order to generate starting structures. We should also note here that 
significant work has already been done in this field (for reviews, see Refs.~\cite{chem_rev_1993_93_2567,rev_comp_chem_1999_13_313,
sado04,wires_comp_mol_sci_2011_1_557}). For example, computational frameworks such as {\sc Open Babel}\cite{j_cheminf_2011_3_33,
j_chem_inf_model_2006_46_991} provide functions which can generate three-dimensional molecular structures from simple {\sc Smiles} 
(Simplified Molecular Input Line Entry System)\cite{j_chem_inf_comput_sci_1988_28_31,j_chem_inf_comput_sci_1989_29_97,
j_chem_inf_comput_sci_1990_30_237} representations. Work along these lines is currently in progress in our laboratory.

In a direct attempt, we represented the ligand sphere as an additional potential in the Kohn--Sham equations. Especially in this case, 
the variational problem is extremely complicated. In order to guarantee the representability of such a potential in terms of an actual 
molecular structure (C-representability), one could try to expand the additional potential in terms of atom-based potentials as introduced by Beratan and 
coworkers\cite{j_am_chem_soc_2006_128_3228}. This technique and other promising optimization techniques are currently explored in our laboratory.
Moreover, we are also considering to employ force-feedback devices for (1) the manual placement of atoms in the chelate-ligand construction 
process and for (2) the manual manipulation of the jacket potential in order to manually guide the optimization of a C-representable jacket potential.
This ansatz is an extension of our direct haptic quantum chemistry approach \cite{marti2009,haag2011,haag2013,haag2014}, in which real-time gradient information is available that
may turn out to be invaluable especially for the first approach towards ligand construction.

\section*{Acknowledgments}

This work has been supported by the Swiss National Science Foundation SNF (project 200020\_144458).

\providecommand{\refin}[1]{\\ \textbf{Referenced in:} #1}

%\bibliographystyle{IntJQuantumChem}
%\bibliography{literature}

\providecommand{\refin}[1]{\\ \textbf{Referenced in:} #1}

\end{document}